# The Financial Heat Machiche: coupling with the present financial crises

Andrei Khrennikov

International Center for Mathematical Modelling
in Physics and Cognitive Sciences,
University of Växjö, S–35195, Sweden,
e–mail: *Andrei.Khrennikov@vxu.se*

We consider dynamics of financial markets as dynamics of expectations and discuss such a dynamics from the point of view of phenomenological thermodynamics. We describe a *financial Carnot cycle* and the financial analogue of a heat machine. We see, that while in physics a perpetuum mobile is absolutely impossible, in economics such mobile may exist under some conditions. Our thermodynamical model for the financial market induces a rather unusual interpretation of the role of financial crises. In contrast to the common point of view, in our model financial crises play a crucial role in functioning of the modern financial market. This is an important (concluding) stage of any financial cycle that is analogous to the stage of cooling in the ordinary Carnot cycle. A financial cycle could not be completed without such a stage as well as the ordinary Carnot cycle. Thus, in spite its destructive (at the first sight) consequences the stage or financial crises is as well important as the stage of "boiling of the financial market" ("heating of expectations").

## 1. Introduction

Financial market is a gigantic system operating with huge ensembles of securities. It is natural to expect that its functioning has common features with the functioning of physical systems, which operate with enormous statistical ensembles. In physics, an ensemble behavior is described by classical statistical mechanics.[1] Therefore, one can expect methods of classical statistical physics, together with methods of classical thermodynamics, to work successfully in financial processes modeling. This model class is the most important component of **econophysics** [2].

---

[1] Ensembles of microparticles, e.g. photons, are described by quantum mechanics. In principle, mathematical methods of quantum mechanics also can be applied to problems in financial sphere [1]. However, in this paper we do not consider quantum financial models, limiting ourselves to those of classical physics.

Chief contribution of the author with the co-authors to econophysics is the realization that a financial market is an **information system**, operating with information values [1]. The key part is played not only by information media, e.g., shares; but also by psychological values, for example, expectations of market players, news and prognoses, that are published in newspapers and announced on TV, politicians' claims and so on.[2] The second most important moment is the realization that for the prediction of the behavior of stock market, one should use collective, rather than individual, variables. In physics, phase-space coordinates of single gas particles are routinely chosen as individual variables. In principle, we can describe the dynamics by means of these variables. Here we have Hamiltonian dynamics or stochastic dynamics of Brownian motion (or of the more general diffusion processes) type. Most financial market models work with this class of variables, application of stochastic processes being the most popular technique, see [2]. Of course, already in physics it has been fruitless to undertake solving a system of Hamiltonian equations for millions of gas molecules.[3] The use of probabilistic description has been proposed. Instead of a system of Hamiltonian equations, we shall use Liouville equation. Describing price fluctuations by means of stochastic processes, one is mostly interested in probabilities dynamics, making use of forward and backward Kolmogorov equations.

Yet there is another approach to the collective variables introduction, that is the **thermodynamical** one. Operating with such variables as **gas temperature, its energy, work**, we can describe the output of the "activity" of big ensembles of gas molecules. We apply this approach to the financial market. It may be said, that this paper is about **financial thermodynamics**, cf. [3--7].

Handling the terms such as the temperature of the market, the energy of the market, work of monetary funds, we create the thermodynamical model of the market. In our model thermodynamical financial variables are defined not only by "hardware" of the market, but by its "software" as well. Not only real economical situation, not just stock prices, but psychological factors as well contribute to, e.g., the temperature of the market [1], [8], [9]. Thus, our model is **informational financial thermodynamics.**

---

[2] Overall, mass media part in our model of functioning of financial market and arising of financial crises is great.

[3] A hundred years ago it was an unsolvable mathematical problem. Nowadays, one can, again in principle, attempt to compute it; however, visualization of millions of trajectories would be still impossible.

Now there is the last important moment. In practically every known econophysical model, financial processes have been considered as objective processes. This point of view is the most distinct in the models based on the theory of stochastic processes. Financial randomness has been considered an objective one [2]. The simplest form of this postulate, see [10], can be found in the financial market model where stock prices considered to be **random walking**. From this point of view, the play strategy better than the one based on coin flipping (heads – go long, tails – go short) cannot be found. Of course, even the advocates of stock market objective randomness understand this model to be somewhat primitive. They tried to "improve" on it by considering new and new classes of stochastic processes: different modifications of Brownian motion, Levy processes, martingales, submartingales, etc., see [2].. However, all these **exercises in probability theory** have been only meant to confirm the postulate of stock dynamics being akin to the stochastic dynamics of gas flow.[4] Well-known fact however is that people for long have known how to make artificial systems to control hot gas flows. A work is performed by these systems; which, meanwhile, consume fuel. Moreover, as the power of the systems and their number grow, the negative influence of their activity becomes more and more considerable.

Financial market functioning studied through thermodynamic analysis bears apparent resemblance to the operation of a heat machine. Thereat, the fact that the machine implementing a financial cycle -- which is a direct analogue of a **Carnot cycle** -- is manmade cannot but strike one's eye. In principle, financial specialists have been monitoring financial cycles for long, and the records can be easily compared to our financial-thermodynamical description, see Malkiel [10].[5] However, econophysics approach makes the structure of financial cycles particularly simple. The analogue to heat

---

[4] This means that modern financial mathematics has a clear ideological dimension. Financial processes are like natural phenomena. One cannot avert a financial hurricane just as one cannot stop a hurricane formation in Caribbean Sea. There are only aftereffects to fight with.

[5] Though the author of this book promotes random walk model, the book contains amazingly demonstrative description of a number of financial Carnot cycles.

machine is the strongest metaphor rendering the shadowy structure given by liberal economics sharp.

The main conclusion of this work is:

After (or, possibly, simultaneously with) the invention of physical heat machine, man contrived financial heat machine. Both physical and informational machines enable work doing (in the latter case the work has the meaning of a profit), realizing cycles of the Carnot type. Unlike the physical heat machine, financial one can break the second law of thermodynamics, allowing, in principle, for creation of financial **perpetuum mobile.**[6] And still, both engines' work feature the last stage, that is steam cooling. Finance notion of this stage is financial crisis. Thus, financial crisis in our model is not generated by the objective randomness of the market. Just as in physical heat machine, it is a critical stage of the working cycle. Without it, physical or financial heat machine just cannot work.

Well, financial crisis is not a financial hurricane of a random nature. It is a man-designed stage of "financial heat machine" operation. Crises elimination would wreck the machine and render financial market unattractive to its creators. Apparently, one should not undertake to fight financial crises as such, but rather alleviate the damage to the "environment". Trying to suppress the Carnot cycle in physics is absurd. No wiser would be to try to forbid the use of the financial Carnot cycle. Informed people will seek to use the cycle nonetheless. World-wide Counteragency for Financial Heat Machines is a non-starter. That is why we should look to lower the "exhaust", and move to more advanced models.

## 2. Carnot cycle at the financial market

Everywhere below price of a share will be denoted by the symbol $p$ and the volume of shares (the number of shares) by the symbol $v$ (may be with some indexes).

---

[6] While the **Maxwell's demon** does not exist in physics, financial Maxwell's demons may exist (and draw good income, too).

For simplicity, let us consider functioning of the financial heat machine for shares of one fixed company. Suppose that totally it was issued $V$ shares. Suppose that a group of people, say $G$, "*designers of a financial heat machine*", was involved in the issue of shares and they got $\Delta V = V - V_1$ shares which they do not sell (for a moment) and the rest of shares $V_1$ circulates at the market.

We discuss the dynamics of the market when the number of shares, which do not belong to $G$, is equal to $V_1$ and the price (also to say open market) is equal to $p_1$. Thus the starting point is characterized by the pair $1 = (V_1, p_1)$.

Then we consider the market dynamics when the expectations of the price grow to $p > p_1$, but (for simplicity) no new shares sold (so people from $G$ just wait and do not sell their shares) and the market arrives at the point $2 = (V_1, p)$. We do not discuss the method how this rise of expectations is achieved.

Then the group $G$ starts to sell shares at the fixed price $p$, and the aim is to sell all their shares, so that, finally, the total volume of shares at the financial market will be equal to $V$. For simplicity we assume that the price of shares during this process will be fixed. Suppose that $G$ sold all its shares. The market arrived at the point $3 = (V, p)$.

Then a *market crash occurs*: the price of shares falls from $p$ to $p_1$. The market arrives at the point $4 = (V, p_1)$.

Then $G$-people buy again $\Delta V$ shares at price $p_1$ reducing volume (at, so to say, "open market") to $V_1$. Thus the market arrives back at our starting point $1 = (p_1, V_1)$.

What will be the result of all these four processes? Namely, the result of the cycle:

1). Rise of the price $1 = (V_1, p_1) \rightarrow 2 = (V_1, p)$.

2). Then $G$-people sell their shares at the high price $2 = (V_1, p) \rightarrow 3 = (V, p)$.

3). Then market crash $3 = (V, p) \rightarrow 4 = (V, p_1)$.

4). Finally, $G$-people buy again shares, but at the low price $4 = (V, p_1) \rightarrow 1 = (V_1, p_1)$.

The result will be the profit:

$$A = \Delta p \, \Delta v,$$

where $\Delta p = p - p_1$, $\Delta v = v - v_1$.

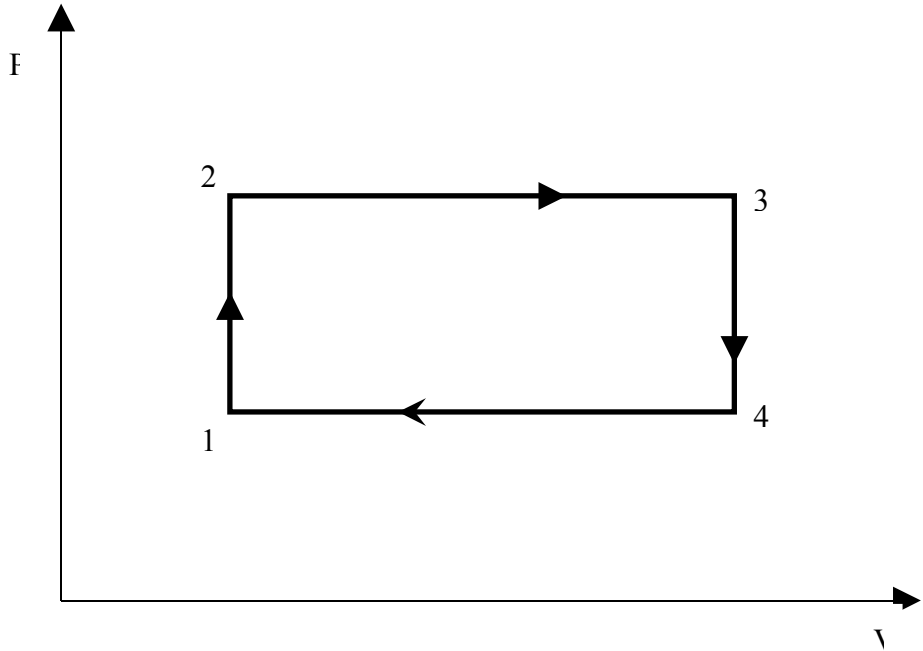

Fig.1: P–V diagram for the financial Carnot cycle

The above described picture is obviously an analog of the Carnot cycle in thermodynamics, which describes say a steam engine.

The **price** $p$ is the analog of **pressure**, and the **volume of the market** (the number of shares) $v$ is the analog of **volume of gas**. The profit $A$ is the analog of **work (or energy)**. If we will continue these physical analogies, **expectations** will be the analog of **vapor**. We mention that Malkiel [10] compared expectations of people at the financial market with Castles-in-the-Air…

By my model if one would be able to **influence the expectations** of the participants of a financial market, then one would be able to make profits using the described financial Carnot cycle.

We remark that the profit produced by the financial Carnot cycle is given by the area of the circuit 1234 (as for the physical Carnot cycle).

## 3. A role of mass-media, a financial perpetuum mobile

To get a better understanding of this concept let us discuss the standard

**steam engine** from a thermodynamic point of view. In a steam engine we have **a boiler, a cooler, and the cylinder**, where the Carnot cycle (in the idealized case) takes place.

The cylinder in our language is the analog of a financial market. What are the analogues of the boiler and the cooler (without which the steam engine will not work)? The boiler is a component which is used to "heat the expectations" of participants of the market and to increase the price. We propose to use *mass media* and other sources of financial influence as the analogue of a boiler. If one will be able to manipulate the expectations of market agents independently of the real situation which corresponds to the activities of the company (to cheat market agents), then this will be the analog of a **perpetuum mobile**. We see, that while in physics a perpetuum mobile is absolutely impossible, in economics such mobile may occur.[1] The role of a cooler will be again played by mass media, which in this case will distribute negative information. The important point here is the possibility of an economic perpetuum mobile for those who can control communications (or for those who control the ones who control communications).
While in real economy, of course, one has to put some fuel into the engine (to produce some real goods and services), the reaction of the stock market and the increase of prices of shares can be considerable higher than changes in real productivity. This shows that the financial market in reality works as an economic perpetuum mobile, where capital is created and destroyed without considerable changes in productivity. This again shows that thermodynamic analogies in economics are incomplete.

Our thermodynamic model for the financial market induces a rather unusual interpretation of the role of financial crises. In contrast to the common point of view, in our model financial crises play the crucial role in the functioning of the modern financial market. This is an important (concluding) stage of a financial cycle that is analogous to the stage of cooling in the ordinary Carnot cycle. A financial cycle could not be completed without such a stage as well as the ordinary Carnot cycle. Thus, in spite its destructive (at the first sight) consequences the stage or financial crises is as well important as the stage of "boiling of the financial market" ("heating of expectations").

## 4. Maxwell demons operating at the financial market

Since in the situation we mentioned above profit is obtained due to using the difference in prices during the market cycle, which means that one sells and buys exactly when it will be profitable. Such an agent works as the Maxwell demon. The mentioned perpetuum mobile will be the perpetuum mobile of the second type (which violates the second law of thermodynamics). The standard objection against the Maxwell demon is that this demon should be in thermal equilibrium with the environment, and therefore it can not use thermal fluctuations of the environment in order to extract energy from the environment: it will itself have thermal fluctuations, and due to these fluctuations the demon will make mistakes, which will equilibrate with a possible gain of energy. But in an economy the Maxwell's demon looks quite possible. If the demon will have a temperature which is sufficiently lower than the temperature of the environment, then a number of mistakes which it makes will be low and the demon will be able to organize a flow of energy from the system. In an economy "temperature", which describes fluctuations, or noise in the system, corresponds to information about the market situation. It is quite possible, that some market agents will be more informed than the others. In this case these agents will be able to use their information to reduce fluctuations in their shares and act as Maxwell demons – to extract profit from the market.

We insist that this shows the way to extract a systematic profit which will not vanish after time averaging. From a physical point of view this corresponds to the fact that each new important information decreases the effective temperature of the market agent, and if one is able to perform the mentioned above Carnot cycle (or, at least, a considerable part of the cycle) before this new information will become common, this market agent will be able to work as a demon of Maxwell and extract systematic profit from the market. Moreover, since the financial market is a complicated nonlinear system, some market players will be able to organize the above mentioned Carnot cycles as a financial market. In this case they will have the needed information about these cycles and therefore will be able to extract profit from the cycles (if the price of organization of the cycle is less than the possible gain in the cycle. Or in other words, if the effectiveness of the financial heat machine is higher than the dissipation).

## 5. More complicated financial Carnot cycle

A primitive financial Carnot cycle presented in section 2 can be modified to exhibit a more complicated behavior (which makes, in particular, less evident its presence at the financial market).

Consider following parameters: volumes of shares $V_1 < V_4 < V_2 < V_3 = V$ and prices $p_1 < p_4 < p_2 < p_3$. We assume that $V_2 - V_1 << V_3 - V_2$ (the first difference is essentially less than the second), $V_3 - V_4 << V_4 - V_1$ and that $p_2 - p_1 << p_3 - p_2$, $p_4 - p_1 << p_3 - p_4$. We now consider a financial Carnot cycle having the same starting point as in section 2. Thus there was totally issued $V = V_3$ shares, at the open market there were sold $V_1$ shares, $G$-people (designers) got $V - V_1$ shares. At the beginning of the process of heating of the positive expectations of this type of shares, the price is equal $p_1$. Cycle:

1). Rise of the price and volume (at the "open market") slowly $1=(V_1, p_1) \to 2=(V_2, p_2)$. [designers of this financial cycle start to sell their shares, the original volume $V_2$, slowly decreases]. Thus the price curve $p_{12}(V)$ increases slowly from $p_1$ to $p_2$. It is defined on relatively small interval, $[V_1, V_2]$.

2). Then very active sale of shares belonging to the financial group $G$ at quickly increasing price $2=(V_2, p_2) \mapsto 3=(V_3, p_3)$. Thus the price curve $p_{23}(V)$ increases rapidly from $p_2$ to $p_3$ on relatively large interval, $[V_2, V_3]$.

3). Then market crash $3=(V_3, p_3) \to 4=(V_4, p_4)$. Thus the price curve $p_{34}(V)$ decreases rapidly from $p_3$ to $p_4$. It is defined on relatively small interval, which is better to denote by $[V_3, V_4]$ (by taking into account cycle's direction). So, $G$-people even buy little bit at such a crashing market, but here is crucial that $V_3 - V_4 << V_4 - V_1$, i.e., this activity does not change essentially their profit.

4). Finally, $G$-people start to buy again shares at the very low price $p_4$: $4=(V_4, p_4) \to 1=(p_1, V_1)$. Here the price curve $p_{41}(V)$ decreases slowly from $p_4$ to $p_1$. It is defined on relatively large interval, which is better to denote by $[V_4, V_1]$ (by taking into account cycle's direction). Now they buy a lot! But it is crucial the difference of the prices: $p_4$ is essentially less than $p_2$ and the price decreases to $p_1$.

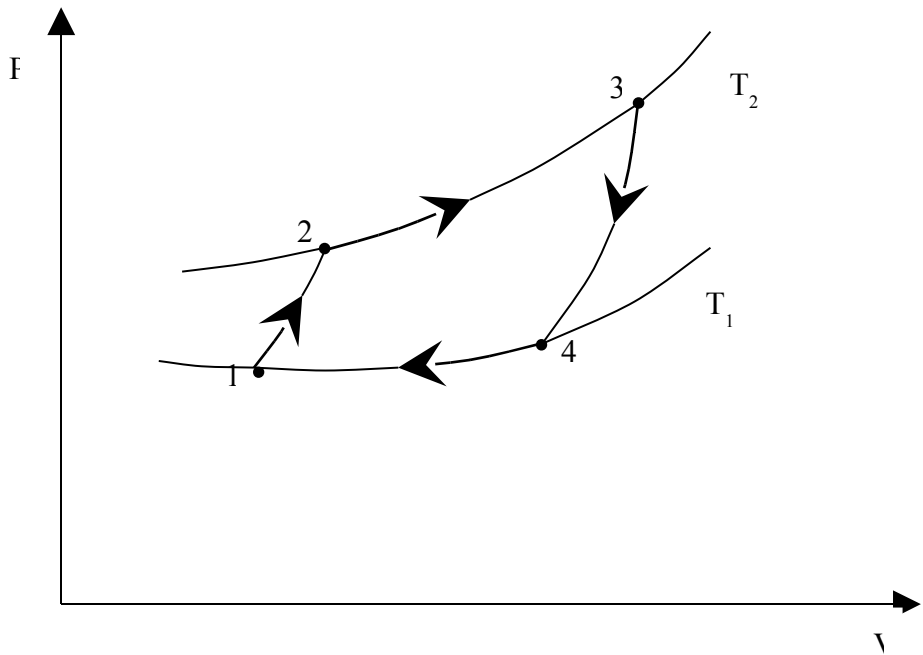

Fig.1A: P–V diagram for the financial Carnot cycle

Curves $p_{ij}(V)$ form a closed graph in the $(V,P)$ plane, the financial Carnot cycle. At the first step of the cycle we get the profit $A_1 = [$area under the graph of $p_{12}(V)]$, then $A_2 = [$area under the graph of $p_{23}(V)]$. After this we spent some money to buy shares for a new cycle: $A_3 = -[$area under the graph of $p_{34}(V)]$, $A_4 = -[$area under the graph of $p_{41}(V)]$. Thus our profit in this more complicated financial Carnot cycle is given by a $A = [$area of the figure of the cycle$]$. Thus, as work in usual thermodynamics, the profit produced by a financial Carnot cycle is given by the area of the circuit 1234.

The financial cycle which was considered in this paper is an analog of the ideal Carnot cycle in thermodynamics. In (in physical as well as financial) reality everything is more complicated. At the real financial market designers of a heat machine need not try to come back to initial point 1, i.e., to prepare a new cycle by buying an essential volume of the same shares. They can escape expenses related to the part 3-4-1 of the cycle. First of all, as it was done in section 2, they can stop any activity during crash. Thus they would not buy shares in the period 3-4. But they can even escape costs of the period 4-1, by, e.g., reemission of shares. I have a personal experience, very negative, with Eriksson's shares, reemission was the last step of Eriksson's Carnot cycle. Instead of reemission, which is profitable in the case of such "respectable corporations" as Eriksson, designers can just start creation of a new financial heat machine and to use profit from the completed cycle with say X-shares to start a new Carnot cycle with say Y-shares. In principle, designers need not be directly involved in the process of emission of shares. They can buy their portion of shares before to start creation of a cycle. In such a case their profits will be less. However, the cost of such an initial operation can be minimized by getting money from banks.

The main deviation of the ideal financial Carnot model from financial reality is consideration of shares of one fixed firm. A powerful financial heat machine (intelligent designers are able to create such machines!) is based on shares of a large group of firms. Creation of large financial heat machines saves a lot of resources. To heat a group of shares is cheaper and sometimes easier than shares of a single firm. Among the most powerful financial heat machines we can mention e.g. Biotechnology -machine or Information-technology-machine, see Malkiel [10] for details. In the case of multi-shares financial heat machine designers buy shares (or participate in issues) of a large number of e.g. biotechnology-firms. Some of these firms have solid grounds, but some of them (if not majority) present really fake projects. Nevertheless, prices of all (at least of majority) biotechnology-shares go up as a result of "fruitful collaboration" with mass-media. After a cycle is done and profits are collected, designers need not to buy (even at very cheap prices) shares of biotechnology –firms. They can start to create e.g. an Information-technology-machine by using a part of profits from the previous heat machine. At the same time it may be profitable to buy shares of selected biotechnology – firms, which seem to do well in future. The greatest financial heat machines have been created at American financial market. The state (which by the way often prosecute creation of small heat machines) may actively participate in creation of huge heat machines, cf. Malkiel [10].

## 6.Comparison of thermodynamic and financial Carnot cycles

In spite of similarity between the thermodynamic Carnot cycle for steam engine and the financial Carnot cycle (and in spite of the fact that I was inspired by physics), it is important to remark that these cycles have different shapes. Analysis of the differences of shapes is up to readers. In any event the explored analogy between price and pressure is not complete, the financial heat machine does its Carnot cycle, but coupling between price and volume differs essentially from coupling between pressure and volume.

Nevertheless, we can try to continue to explore econophysical metaphor and consider curves 2-3 and 4-1 as corresponding to fixed financial temperatures. Here $T_1 \ll T_2$. Financial temperature increases drastically during period 1-2 development of the market, then it is constant $T=T_2$ (at least in the ideal model) during period 2-3. The market is very hot, sufficiently hot. Then it goes down drastically during period 3-4, and finally, it is constant again, but temperature is very low $T=T_1$.

We remark once again that the presented econophysical model is based on a metaphor of classical thermodynamics -- in contract to majority of known econophysical models on dynamics of expectations at the financial market, which were based on the use of quantum-like metaphors, see [1], [11-16].

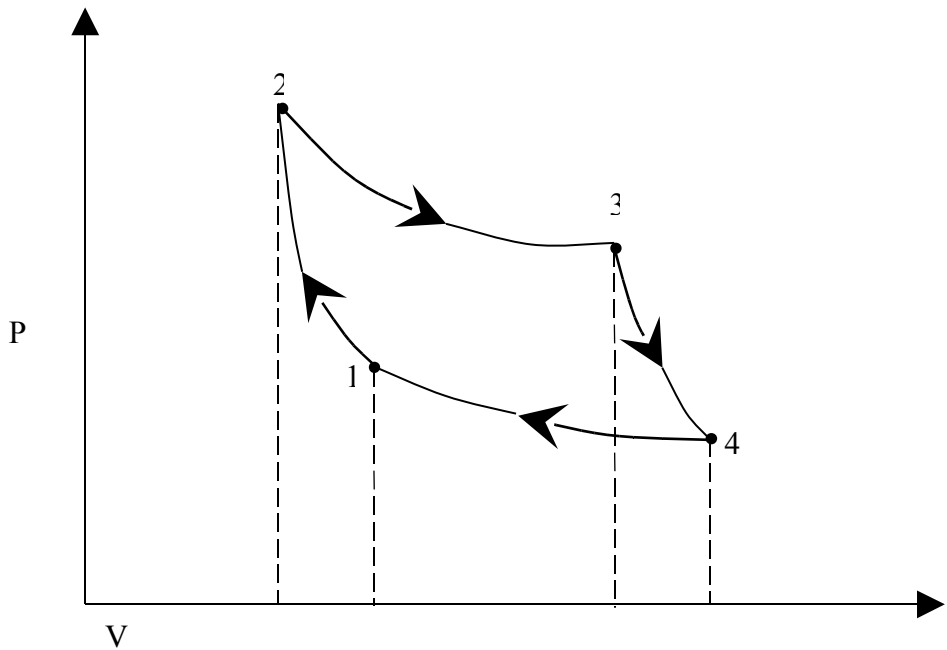

Fig.2: P–V diagram for the physical Carnot cycle

**Conclusions:** 1). Majority of financial crises (including world's crises) are not catastrophic events induced by uncontrollable randomness of the complex financial system, the modern financial market. They are simply important phases of functioning of financial heat machines created by the human intellect.

2). Surprisingly, our thermodynamic model matches with the conventional model of the financial market based on the efficient market hypothesis. In our model, as in the conventional model, there is no possibility of arbitrage for the great majority of actors of the financial market. The only difference is that by our model such a possibility exists for small groups of actors -- designers of financial heat machine. However, from the viewpoint of the conventional model which is based on the measure-theoretic approach (due to Kolmogorov), the measure of the set of those actors-designers is equal to zero. So, from the measure-theoretic viewpoint one can neglect by this set. However, we demonstrated that, although in a purely mathematical model, one can neglect by a group of zero measure, it cannot be done at the real financial market.

Of course, in a "fair" economic environment there should be a correspondence between the market expectations and the situation describing the activities of the company. In our language this means that one has to put "fuel" into the boiler. However, nowadays expectations depend only indirectly on the situation in real economy. Therefore our in our model coupling with real economy is totally neglected. Recently I presented my model at the seminar of the Department of Economics of Växjö University. I was strongly criticized by prof. Jan Ekberg who emphasized the role of real economics. He pointed out that profits produced by a financial Carnot cycle are based on work of people in real economics and, in fact, their wok is the real fuel of financial Carnot cycles. The mass-media does not produce fuel for financial heat machines. It just pumps mentioned fuel (the result of work of real economics) to these machines. In any event the presented model should be completed with functioning of productive economics. It will be done in a coming publication.

I would like to thank T. Nieuwenhuizen for some stimulating discussions on the foundations of thermodynamics and especially the second law of thermodynamics, S. Kozyrev for discussions on analogy between theory of disordered systems and economy, E. Haven for knowledge transfer on the financial market and econophysics.